\documentclass[twocolumn,showpacs,amsmath,amssymb,superscriptaddress]{revtex4}

\usepackage{graphicx}
\usepackage{dcolumn}
\usepackage{bm}

\begin{document}

\title{Stability of hydrogenated group-IV nanostructures: magic structures of diamond nanocrystals and Silicon quantum dots}

\author{Xiaobao Yang}
\email[corresponding author: ]{scxbyang@scut.edu.cn}
\affiliation{Department of Physics, South China University of
Technology, Guangzhou 510640, People's Republic of China}
\author{Hu Xu}
\affiliation{Center of Super-Diamond and Advanced Films (COSDAF)
Department of Physics and Materials Science, City University of Hong
Kong, Hong Kong SAR, China}
\author{Yu-Jun Zhao}
\affiliation{Department of Physics, South China University of
Technology, Guangzhou 510640, People's Republic of China}
\author{Boris I. Yakobson}
\affiliation{Department of Mechanical Engineering and Materials
Science, Department of Chemistry, and The Richard E. Smalley
Institute for Nanoscale Science and Technology, Rice UniVersity,
Houston, Texas 77005, USA}

\date{\today}

\begin{abstract}

We have developed an effective model to investigate the energetic
stability of hydrogenated group-IV nanostructures, followed by
validations from first-principles calculations. It is found that the
Hamiltonian of X$_{m}$H$_{n}$ (X=C, Si, Ge and Sn) can be expressed
analytically by a linear combination of the atom numbers ($m$, $n$),
indicating a dominating contribution of X$-$X and X$-$H local
interactions. As a result, we explain the stable nanostructures
observed experimentally, and provide a reliable and efficient
technique of searching the magic structures of diamond
nanocrystals(Dia-NCs) and Silicon quantum dots(SiQDs).

\end{abstract}


\maketitle

Semiconductor nanocrystals have been greatly attractive and
intensively investigated\cite{Alivisatos1996,Burda2005}. These
nanomaterials extend the physics of reduced dimensions and offer the
opportunity for fundamental study of the regime between
nanostructure and bulk states\cite{Reimann2002}, which have also
brought wide applications as nanoscale electronic and optical
devices\cite{Pavesi2000}, fluorescent biological
labels\cite{Bruchez1998}, quantum computation media\cite{Loss1998}
etc. Besides II-VI and III-V semiconductor compounds\cite{Peng2000},
various hydrogenated group-IV (C\cite{Dahl2003,Landt2009},
Si\cite{Buuren1998,Belomoin2002}, and Ge\cite{Bostedt2004})
nanocrystals with $sp^3$ hybridizations have been synthesized and isolated, whose optical
response depends on particle size, shape and symmetry. Theoretical
studies\cite{Rohl1998,Vasi2001,vr2009},  have focused on optical
properties of these nanocrystals and employed various methods for
accurate calculations of adsorption spectrum.

The determination of stable structures, which dominate the optical
properties of nanocrystals\cite{Dahl2003,Landt2009}, is not well
understood so far. Previous calculations constructed
Dia-NCs\cite{vr2009} according to the synthesis experiments, which
indicated that larger members of the series have smaller
surface-to-volume ratios and lower hydrogen-to-carbon
ratios\cite{Dahl2003}. According to the Wulff energy, the
polycrystalline wire of five-fold symmetry is more stable than
single-crystal types\cite{zhao2003}, for silicon nanowires with the
diameter less than 6 nm. Recent studies\cite{chan2006,lu2007}
searched the magic structures of silicon nanowires using genetic
algorithm, in which the energies are calculated with classical
potential in Hansel-Vogel (HV) formulism to reduce the time cost.

It is challenging to determine the stable configuration from numerous
possible candidates and there are three main obstacles for searching
magic structures of X$_m$H$_n$ (X=C, Si, Ge and Sn): 1) the accurate
calculation of the total energy is necessary but often
computationally expensive; 2) over many isomeric structures should
be considered and the number of these structures increases sharply
as $m$ and $n$ increase; 3) the minimum of $n$ is not clear for a
certain $m$, though the maximum of $n$ is trivially $2m+2$.

In this letter, we investigate hydrogenated group-IV nanocrystals by both model analysis and first-principles
approaches. We proposed an effective model and gave an analytical
expression of the Hamiltonian for X$_m$H$_n$ (X=C, Si, Ge and Sn)
with the numbers of atoms $(m, n)$, as is confirmed by the
first-principles calculations. Our finding provides an efficient and
reliable avenue of searching magic structures of Dia-NCs and SiQDs,
which extends our understanding on experiment observed stable
nanocrystals.

In our model, we assume the Hamiltonian of X$_m$H$_n$ (X=C, Si, Ge
and Sn) to be as
\begin{equation}
{\cal H}=\sum_{i=1}^{m}({\cal H}_{\textrm{int}}+{\cal H}_0)-m\mu_{\textrm{X}}-n\mu_{\textrm{H}}
\end{equation}
where ${\cal H}_{\textrm{int}}$ and ${\cal H}_0$ are the contributions from interactions and self energies,
and $\mu_{\textrm{X}}$($\mu_{\textrm{H}}$) is the
chemical potential for X(H) atom. The sum runs over all the group-IV atoms and we have ${\cal H}_0=\mu_{\textrm{X}_\textrm{0}}+p_{i}\mu_{\textrm{H}_\textrm{0}}$,
where $p_i$ is the number of H atoms in saturated group of the $i$th X atom and $\mu_{\textrm{X}_\textrm{0}}$($\mu_{\textrm{H}_\textrm{0}}$) is the
isolated atomic energy for X(H) atom. ${\cal H}_{\textrm{int}}$ includes the
energy contributions from bonded X--atom pair($-E_{\textrm{X--X}}$) and the saturated
group of the $i$th X atom($-E_{\textrm{--XH}_{p_{i}}}$). As shown in the inset of Fig.1, every X atom has four nearest neighbors and
every X$-$X bond is shared by two X atoms. This leads the energy
contribution corresponding to the $i$th X atom to be $-2E_{\textrm{X--X}}$ for $p_i=0$, and $-1.5E_{\textrm{X--X}}-E_{-\textrm{XH}}$ for $p_i=1$
analogically.  Thus, ${\cal H}_{\textrm{int}}=-(2-0.5p_i)E_{\textrm{X--X}}-E_{\textrm{--XH}_{p_{i}}}$.

We assume that the interaction between X and H atoms is localized and thus $E_{\textrm{--XH}_{p_{i}}}=p_{i}E_{\textrm{X--H}}$, $p_i=0,1,2,3$.
With $n=\sum_{i=1}^{m}p_i$, we found that Eq. (1) can be written
as ${\cal H}=am+bn$, where $a=(-2E_{\textrm{X-X}}-\mu_{\textrm{X}}+\mu_{\textrm{X}_\textrm{0}})$ and
$b=(-E_{\textrm{X--H}}+0.5E_{\textrm{X--X}}-\mu_{\textrm{H}}+\mu_{\textrm{H}_\textrm{0}})$.
This implies that the energy of X$_m$H$_n$ is determined by
the interaction strength of X$-$X and X$-$H, chemical potentials, and
the atom numbers $(m, n)$.

Instead of scanning the parameter space (a, b)\cite{yang2003}, the Simplex method\cite{brandt,kanamori1966,stolze1980} is efficient to determine the combinations of lowest energy
from a set of possible integer combinations $(m, n)$, which indicates
that the stable configurations correspond to corners, edges and faces\cite{stolze1980}. We will obtain the ground states if we construct enough
restricting inequalities and solve the corresponding linear
euquations\cite{stolze1980}. In our case, we have $m\geq1$ and $n\leq2m+2$. The key task is to
determine the lower limit of $n$ for a certain $m$, for which it is
difficult to find out the expected inequalities. We search the least
$n$ as follows\cite{recon}:  i ) starting from one of the stochastic configuration
of X$m$H$n$ ($n\leq2m+2$), exchange X and H atoms and saturate the configuration
with H atoms when necessary; ii ) we accept the new configuration
when the H atom number is non-increasing, otherwise the new geometry
will be accepted with the probability of  $1/|dn|$, where $dn$ is the increment
of H-atom number. We find that $n$ will converge into the minimum
after hundreds of iterations.

\begin{figure}[tdb]
\includegraphics[width=0.45\textwidth,trim=0 0 0 0,clip] {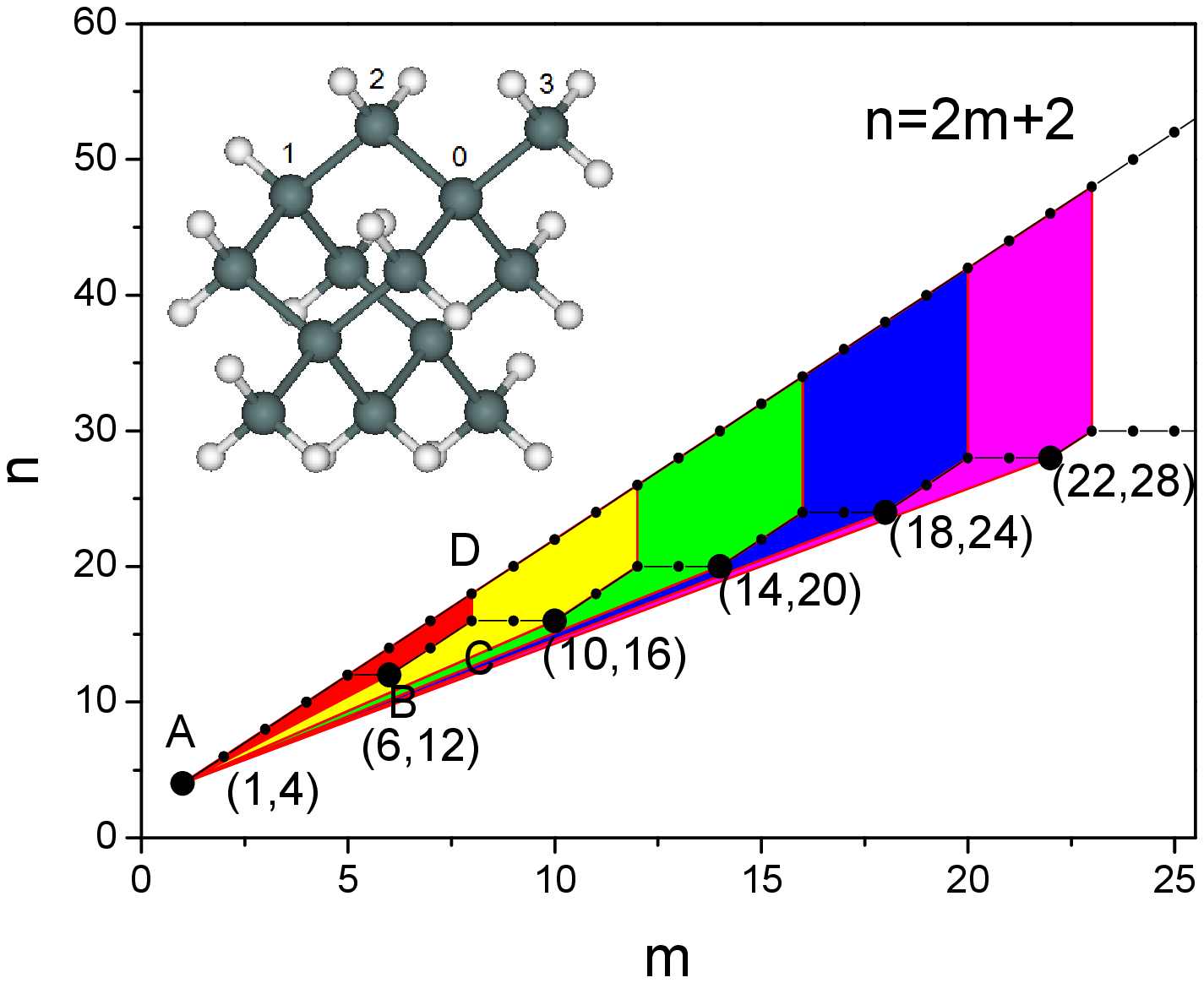}
\caption{\label{fig1} (Color online) Possible ($m$, $n$) for
X$_{m}$H$_{n}$. The convex corresponds to the local minimum and
magic structures of hydrogenated nanostructures. }
\end{figure}

Figure 1 shows the upper and lower
limit of $n$ as a function of $m$. It is not a standard Simplex because
both $m$ and $n$ will be increasing with an increasing nanocrystal size
of X$_m$H$_n$. However, we will obtain a convex quadrangle ABCD (colored
in red) if a restriction of $m\leq8$ is considered. The convex A (1, 4) and
B (6, 12) are corresponding to stable configurations, while C (8,
16) and D (8, 18) are not since they are induced by the artificial
restriction of $m\leq8$. Thus, XH$_4$ and X$_6$H$_{12}$ will be stable configurations
for group-IV nanocrystals. Analogically, we will obtain a new convex
quadrangle (color in red and yellow) and find another stable
configuration of X$_{10}$H$_{16}$ in place of X$_6$H$_{12}$, if we consider a
restriction of $m\leq12$. Besides, we find that X$_{14}$H$_{20}$, X$_{18}$H$_{24}$ and X$_{22}$H$_{28}$
are also stable configurations. It should be noted that, except for
XH$_4$, all other nanocrystals are meta-stable states, since they are
local convex ascribed to the size confinement. As is known, the size
of nanocrystals increases with increasing reaction time as more
material is added to the surfaces.

Till now, we have settled the main obstacles in searching magic
structures of group-IV nanostructures according to our model
analysis. Firstly, we can estimate the Hamiltonian simply by
$E_{\textrm{tot}}=m\mu_{\textrm{X}_\textrm{0}}+n\mu_{\textrm{H}_\textrm{0}}-(2mE_{\textrm{X--X}}-0.5nE_{\textrm{X--X}}+nE_{\textrm{X--H}})$,
since the total energies ($E_{\textrm{tot}}$) directly obtained from the
first-principles calculations do not involve the environment-related
chemical potentials and in fact correspond to the Hamiltonian in Eq.
(1) with $\mu_{\textrm{X}}=\mu_{\textrm{H}}=0$. Secondly, we should only consider few possible
candidates of X$_m$H$_n$ with various $n$ for a certain $m$, as isomeric
structures with the same chemical formula will possess comparable
total energies. Thirdly, we have searched the least $n$ corresponding
to a certain $m$ iteratively and determined magic structures by the
Simplex method.

\begin{figure}
\includegraphics[width=0.45\textwidth,trim=0 0 0 0,clip] {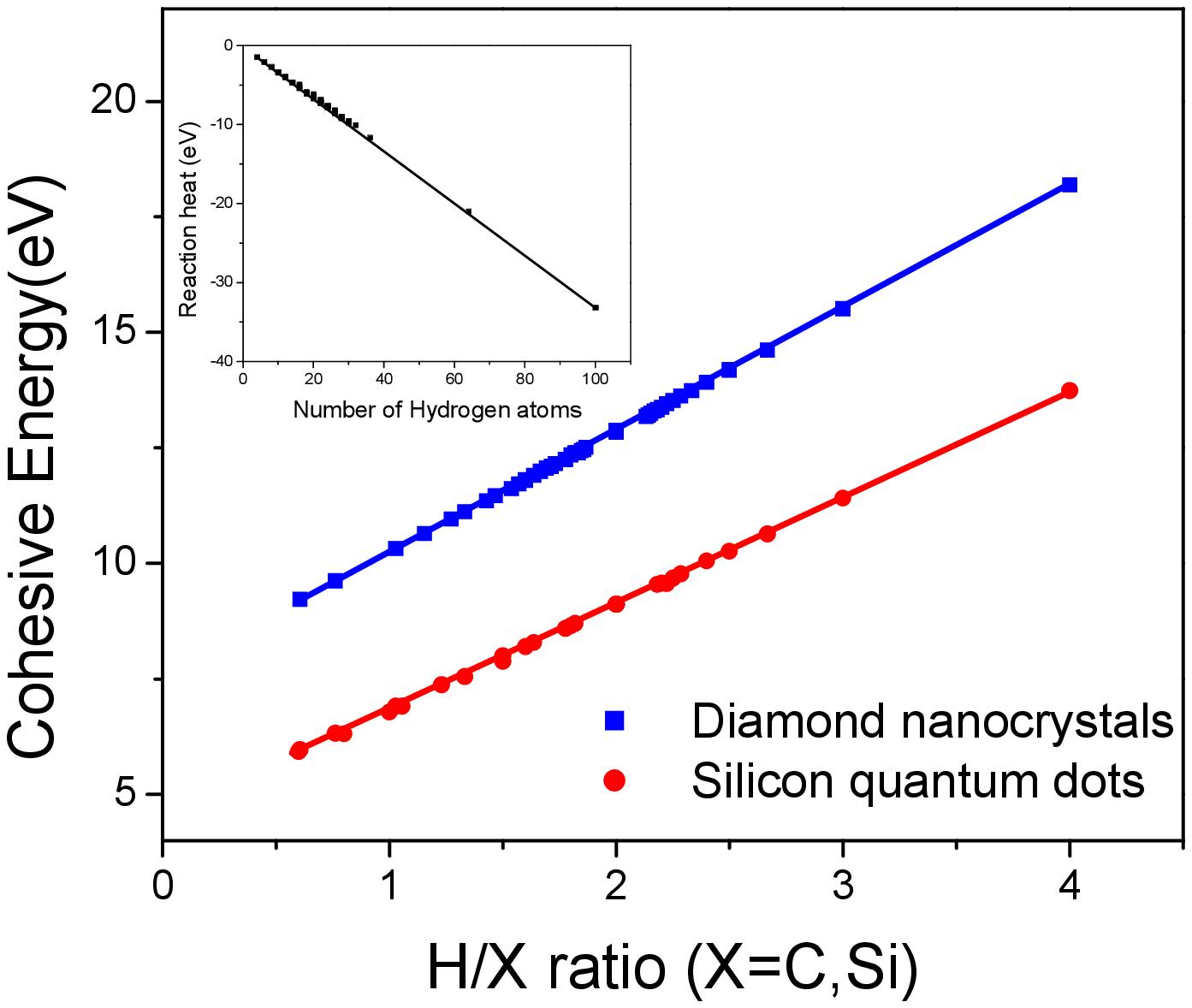}
\caption{\label{fig1} (Color online) Cohesive energies of Dia-NCs and SiQDs as a function of the H/(C, Si)
ratio. Inset shows the linear dependence between the reaction heat
of Dia-NCs and hydrogen numbers.}
\end{figure}

To verify the reliability of the above model, we
investigate the energetic stability of group-IV nanoctrystals (with
the example of Dia-NCs and SiQDs) to search the magic structures,
using the first -principles method implemented in Vienna Ab initio
Simulation Package (VASP)\cite{kresse1993,kresse1996}.   We use Vanderbilt ultrasoft
pseudopotentials\cite{vander1990} and the exchange correlation with the generalized
gradient approximation given by Perdew and Wang\cite{Perdew1992}. We set the
plane-wave cutoff energy to be 350 eV and the convergence of the
force on each atom to be less than 0.01 eV/$\dot{\textrm{A}}$   mesh of $\textbf{k}$ space is
used and the vacuum distance is set to be 9 $\dot{\textrm{A}}$, which is enough to
make the systems isolated.

We define the cohesive energy ($E_{\textrm{coh}}$) per X
atom in the nanocrystals X$_m$H$_n$ as $E_{\textrm{coh}}=(m\mu_{\textrm{X}_\textrm{0}}+n\mu_{\textrm{H}_\textrm{0}}-E_{\textrm{tot}})/m$,
i.e., $E_{\textrm{coh}}=2E_{\textrm{X--X}}-(0.5E_{\textrm{X--X}}-E_{\textrm{X--H}})\alpha$, with $\alpha$ is the H/X ratio
$(n/m)$. Figure 2 shows the cohesive energy of X$_m$H$_n$ (X=C, Si) as a
function of the H/X ratio. As predicted, the cohesive energy
decreases with the decrease of the H/X ratio following a linear
relationship, approaching the value of bulk material ($-\mu_{\textrm{Diamond-bulk}}$=7.65eV and $-\mu_{\textrm{Si-bulk}}$=4.67eV\cite{yang2005}) when
the H/X ratio reaches zero. In addition, we can make a deduction
that the reaction heat $Q=E_{\textrm{tot}}(\textrm{C}_m\textrm{H}_n)-E_{\textrm{tot}}(\textrm{C}_m,\textrm{Diamond})-E_{\textrm{tot}}(\textrm{H}_n,\textrm{H}_2)
=n(0.5E_{\textrm{C--C}}-E_{\textrm{C--H}}+0.5E_{\textrm{H--H}})$ will also have a linear dependence on the H
atom number $n$, confirmed by our calculations (shown in the inset of
Fig. 2). Thus, the interaction in X$_m$H$_n$ is dominated by the localized
X$-$X and X$-$H interactions and the Hamiltonian estimated in the model
by $E_{\textrm{tot}}=m\mu_{\textrm{X}_\textrm{0}}+n\mu_{\textrm{H}_\textrm{0}}-(2mE_{\textrm{X--X}}-0.5nE_{\textrm{X--X}}+nE_{\textrm{X--H}})$ is at the accuracy level of the first-principles approach.

\begin{figure}
\includegraphics[width=0.45\textwidth,trim=0 0 0 0,clip] {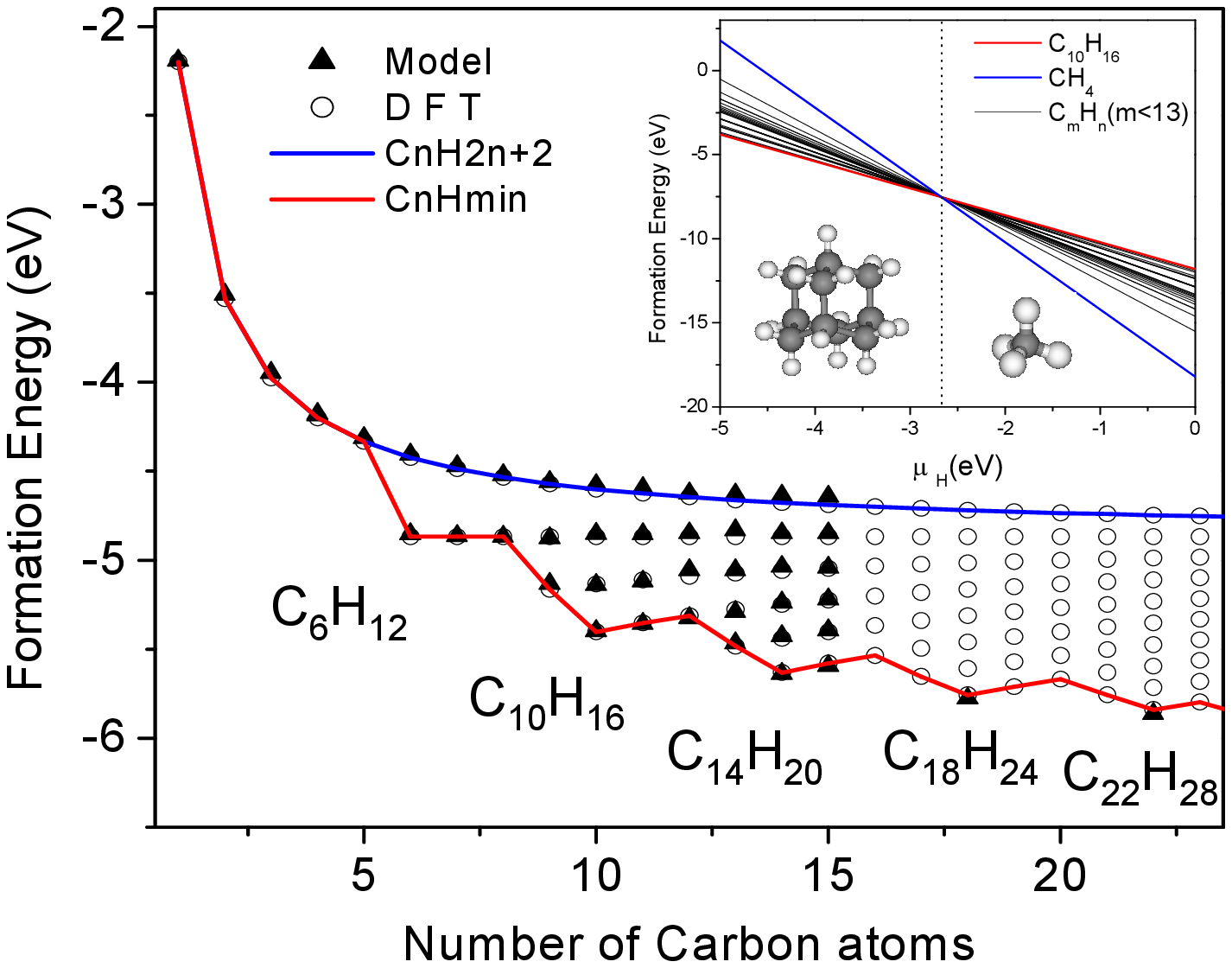}
\caption{\label{fig1} (Color online) Formation energies as a
function of the number of Carbon atoms for diamond nanocrystals with $\mu_{\textrm{H}}=-4$eV
. Inset shows the chemical phase diagram for C$_m$H$_n$ with $m\leq12$.}
\end{figure}

In the following, we investigate the magic structures of Dia-NCs
exemplified by C$_m$H$_n$. We calculated the total energies of
nanocrystals with carbon atoms $m\leq12$ and obtained the formation energies
as a function of hydrogen chemical potential by $E_f=(E_{\textrm{tot}}-m\mu_{\textrm{X}_\textrm{0}}-n\mu_{\textrm{H}_\textrm{0}}-n\mu_{\textrm{H}})/m$. As shown in the
inset of Fig.3, we find that there is a critical point $\mu_{0}$($\mu_{0}=0.5E_{\textrm{C--C}}-E_{\textrm{C--H}}\simeq-2.63\textrm{eV}$) of chemical
potential, at which the formation energies are the same for all
these nanocrystals. Below the critical point C$_{10}$H$_{16}$ is the most
stable, while CH$_4$ is the most stable when $\mu_{\textrm{H}}$ is above the critical
point. As predicted, CH$_4$ and C$_{10}$H$_{16}$ are stable states when the
number of carbon atoms $m\leq12$.

 Figure 3 shows the formation energies for
various C$_m$H$_n$ with the chemical potential of hydrogen $\mu_{\textrm{H}}=-4$eV. The hollow
circles are from our model prediction, which is in excellent
agreement with the ones from the first-principles calculations
(marked with solid triangles). It is noted that the isomeric
structures have similar formation energies especially for the
chemical potential far away from the critical point, though there
are remarkable difference in their total energies. For a certain $m$,
the formation energy decreases as the hydrogen atom number $n$
decreases. The magic structures can be found at the local minimum of
the formation energies, such as C$_{10}$H$_{16}$, C$_{14}$H$_{20}$, C$_{18}$H$_{24}$ and C$_{22}$H$_{28}$.
All these structures are consistent with the previous model
analysis, which have also been confirmed by experiments\cite{Dahl2003,Landt2009}. We can
also predict that, for the chemical potential $\mu_{\textrm{H}}>\mu_{\textrm{0}}$, the formation
energy will decreases as the hydrogen atom number $n$ increases for a
certain $m$, thus the stable structures are alkane C$_m$H$_{2m+2}$.

For hydrogenated group-IV nanocrystals, magic structures often
correspond to the ones with the least or most hydrogen atoms,
according to the linear dependence of total energies on H/X ratio.
In our previous study\cite{xu2009}, we have investigated magic structures of
hydrogenated SiQDs, which are in agreement with the ones of Dia-NCs.
It is reasonable as our model analysis showed that the magic
structures of X$_m$H$_n$ are the same for all group-IV elements. So far,
we have not considered any possible reconstructions, which might
induce strain and instability, especially for Dia-NCs with small
size. For SiQDs, however, the formation of dimers is common and
often dominates the surface reconstruction\cite{Belomoin2002,zhao2003,puzder2003}, which will further
decrease the hydrogen number. According to our calculations, we find
that the $E_{\textrm{coh}}$ of SiQDs with reconstructions also follow the linear
dependence on the H/Si ratio.

\begin{figure}
\includegraphics[width=0.45\textwidth,trim=0 0 0 0,clip] {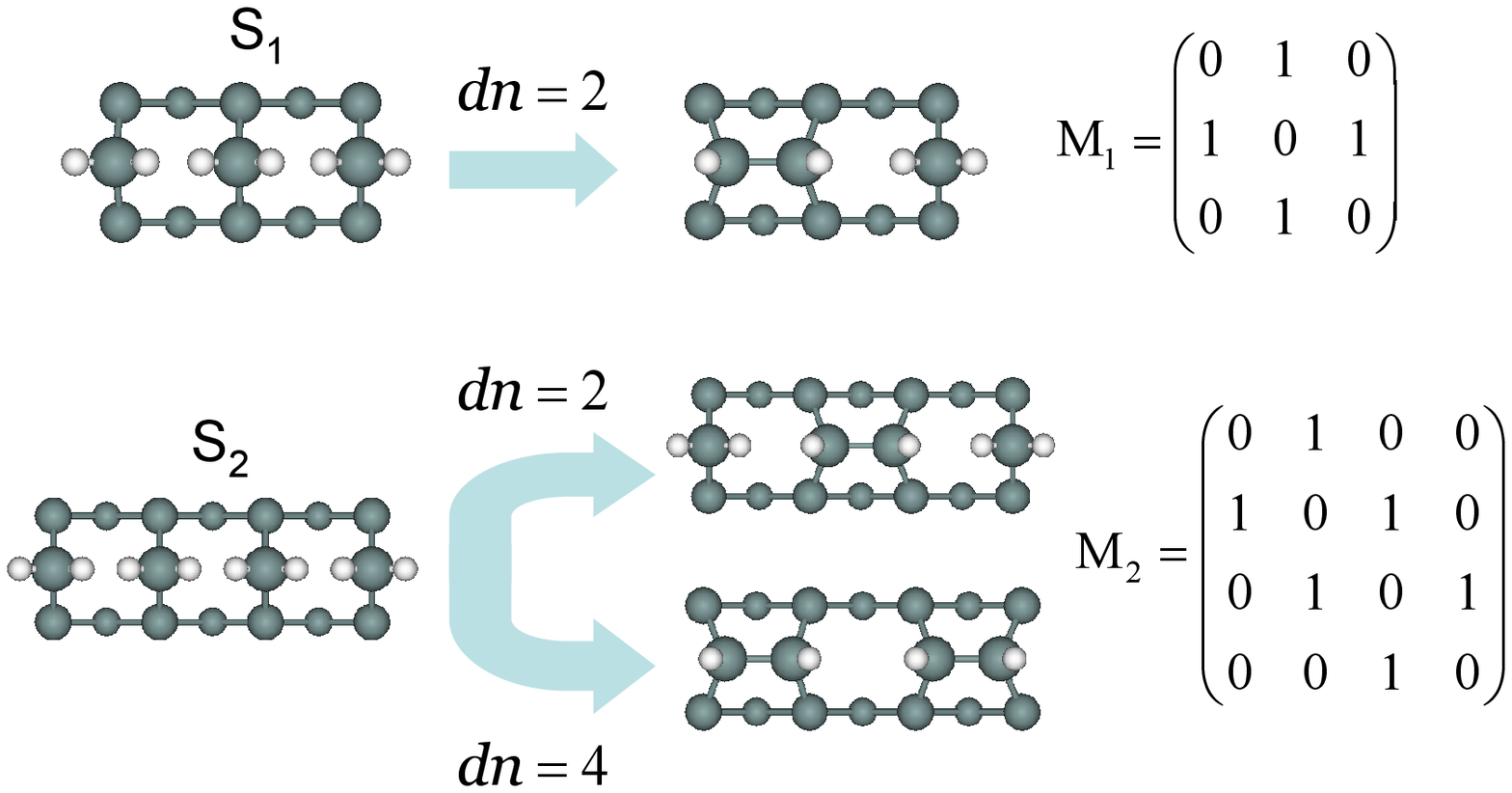}
\caption{\label{fig1} (Color online) The possible dimer
reconstructions and their effect on the number of hydrogen atoms..}
\end{figure}

\begin{figure}
\includegraphics[width=0.45\textwidth,trim=0 0 0 0,clip] {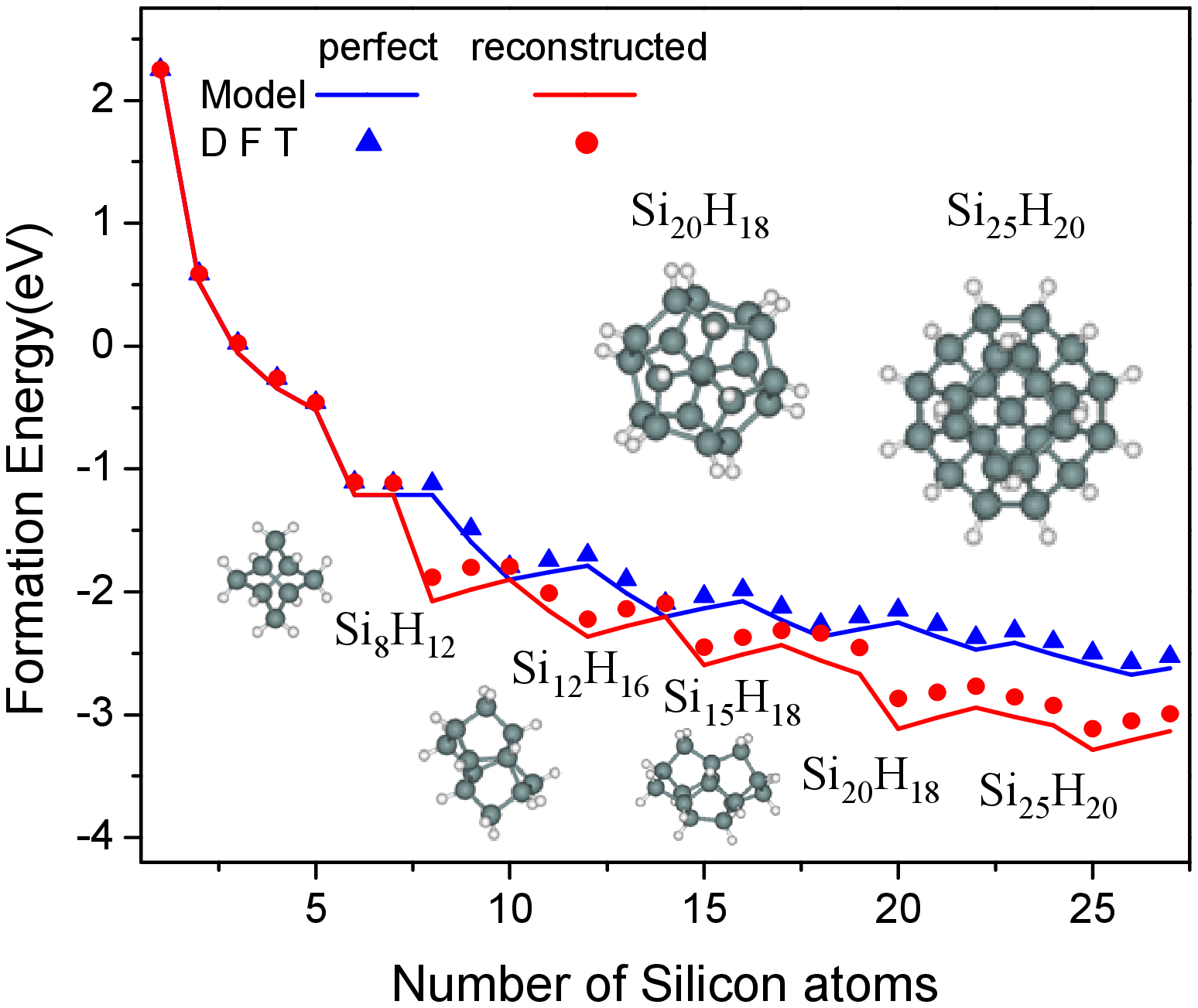}
\caption{\label{fig1} (Color online) (Color online) Formation
energies as a function of the number of Silicon atoms for SiQDs with and without the dimer reconstructions.}
\end{figure}

For simplicity, we consider a dimer reconstruction on
(100) facet as is shown in Fig. 4. The hydrogen number of structure
S$_1$ will decrease by 2 when the dimer forms. For structure S$_2$, the
hydrogen number will decrease by 2 or 4, depending on the selection
of Si atoms for reconstructions. To find out the max decrement of
hydrogen number, we construct a matrix comprised of 0 and 1
according to the arrangement of Si atoms with $-$SiH$_2$: the element
M$(i,j)$ in the matrix is 1 when a dimer could be formed by the $i$th
and $j$th Si atoms; otherwise, it is 0. M$_1$ and M$_2$ correspond to the
structures S$_1$ and S$_2$ shown Fig. 4. The max decrement of hydrogen
number is equal to the rank of the matrix. It can be comprehensible
that the max decrement of hydrogen number is independent of the
order of Si atoms, and the rank of matrix is an intrinsic parameter
that conserves in the elementary transformations. For SiQDs with
various facets, we can construct a block matrix:M=(M$_1$,0,0;0,M$_2$,0;0,0,...). The max decrement
of hydrogen number should be calculated as:rank(M)=rank(M$_1$)+rank(M$_2$)+... We search the least
hydrogen number for SiQDs with a certain number $m$ through the
similar procedure.

Figure 5 shows the formation energies of Si quantum dots with and
without dimmer reconstructions with $\mu_{\textrm{H}}=-4$eV. The dimer reconstruction
decreases the formation energies and changes the magic structures.
For example, Si$_{10}$H$_{16}$ might not be the magic structure as its
formation energy is higher than that of Si$_8$H$_{12}$, which can be
obtained from Si$_8$H$_{16}$ with two dimer reconstructions. Besides,
Si$_{12}$H$_{16}$, Si$_{15}$H$_{18}$, Si$_{20}$H$_{18}$, and Si$_{25}$H$_{20}$ are new magic structures. The
reconstructions also have significant effect on the symmetry of
SiQDs. Stable SiQDs without reconstructions tend to be octahedron
enclosed by (111) facet, e.g. Si$_{35}$H$_{36}$. However, the reconstructed
Si$_{20}$H$_{18}$ has the symmetry of $C_{3v}$ and Si$_{25}$H$_{20}$ is approximately
spherical.

In summary, we have proposed an effective method for investigating
the stability of H-terminated group-IV nanostructures. We express
the Hamiltonian of X$_m$H$_n$ (X=C, Si, Ge and Sn) into the linear
combination of the atom numbers and obtained the total energies with
high reliability. We find that the isomeric structures are nearly
energetically degenerated and the stable structures correspond to
the ones with the least or most hydrogen atoms. The predicted stable
Dia-NCs are in good agreement with experimental observations. With
the overcome of the bottlenecks for stability determination, our
model provides an efficient technique of searching magic structures.

\begin{acknowledgments}
This work was supported by NSFC under Grant No. 10704025, and the
New Century Excellent Talents Program (Grant No. NCET-08-0202).
\end{acknowledgments}


\begin{thebibliography}{99}
\bibitem {recon} For simplicity, we did not consider any reconstruction.
\end{thebibliography}
\end{document}